\documentclass[useAMS,usenatbib]{mn2e}
\usepackage{epsfig}
\usepackage{graphicx}
\usepackage{amsmath}

\newcommand{\ba}{\begin{eqnarray}}
\newcommand{\ea}{\end{eqnarray}}
\newcommand{\be}{\begin{equation}}
\newcommand{\ee}{\end{equation}}

\def\go{\mathrel{\raise.3ex\hbox{$>$}\mkern-14mu
             \lower0.6ex\hbox{$\sim$}}}
\def\lo{\mathrel{\raise.3ex\hbox{$<$}\mkern-14mu
             \lower0.6ex\hbox{$\sim$}}}

\begin{document}

\title[Inertial-acoustic Oscillations of Black-Hole Accretion Discs]
{Inertial-Acoustic Oscillations of Black-Hole Accretion Discs
with Large-Scale Poloidal Magnetic Fields}
\author[C. Yu and D. Lai]
{Cong Yu$^{1,2,3}$\thanks{Email:cyu@ynao.ac.cn},
and Dong Lai$^2$\\
$^1$ Yunnan Observatories, Chinese Academy of Sciences, Kunming, 650011, China \\
$^2$ Department of Astronomy, Cornell University, Ithaca, NY
14853, USA \\
$^3$ Key Laboratory for the Structure and Evolution of Celestial
Objects, Chinese Academy of Sciences, Kunming, 650011, China}

\pagerange{\pageref{firstpage}--\pageref{lastpage}} \pubyear{2015}

\label{firstpage}
\maketitle

\begin{abstract}
We study the effect of large-scale magnetic fields on the
non-axisymmetric inertial-acoustic modes (also called p-modes)
trapped in the innermost regions of accretion discs around black
holes (BHs). These global modes could provide an explanation for
the high-frequency quasi-periodic oscillations (HFQPOs) observed
in BH X-ray binaries. There may be observational evidence for the
presence of such large-scale magnetic fields in the disks since
episodic jets are observed in the same spectral state when HFQPOs
are detected. We find that a large-scale poloidal magnetic field
can enhance the corotational instability and increase the growth
rate of the purely hydrodynamic overstable p-modes. In addition,
we show that the frequencies of these overstable p-modes could be
further reduced by such magnetic fields, making them agree better
with observations.
\end{abstract}


\section{Introduction}


Black-hole (BH) X-ray binaries exhibit a wide range of
variabilities (e.g. van der Klis 2006; Remillard \& McClintock 2006;
Yu \& Zhang 2013)
Of particular interest is the High-Frequency Quasi-Periodic
Oscillations (HFQPOs; Remillard \& McClintock 2006; Belloni et al.~2012;
Belloni \& Stella 2014).
They have frequencies (several tens to a few hundreds Hz)
comparable to the orbital frequency at the Innermost Stable
Circular Orbit (ISCO) around the BH (with mass $M\sim 10
M_{\odot}$) and thus provide a unique probe to study accretion
flows near BHs and the effects of strong gravity. These HFQPOs are
only observed in the intermediate spectral state of BH X-ray
binaries, when the system transitions between the low/hard state
(with the X-ray emission dominated by power-law hard photons) and
the thermal state (with the emission dominated by thermal disk
photons). Interestingly, it is during the intermediate state when
episodic jets are observed from the BH X-ray binaries (e.g. Fender
et al.~2004).
Currently, the physical origin of HFQPOs remains elusive, and a
number of ideas and models have been suggested or explored,
including the orbital motion of hot spots in the disc (Stella et
al. 1999; Schnittman \& Bertschinger 2004; Wellons et al. 2014),
nonlinear resonances (Abramowicz \& Kluzniak 2001; Abramowicz et
al. 2007), and oscillations of finite accretion tori (Rezzolla et
al. 2003; Blaes et al. 2006). A large class of models are based on
BH discoseismology, in which HFQPOs are identified as global
oscillations modes of the inner accretion discs (see Kato 2001,
Ortega-Rodriguez et al. 2008 and Lai et al.~2013 for reviews).

In this paper, we study the effect of large-scale magnetic fields
on non-axisymmetric ($m>0$) p-modes [also called inertial-acoustic
modes; see Kato (2001) and Wagoner (2008) for reviews] trapped in
the innermost region of the accretion disc around a BH. Lai \&
Tsang (2009) and Tsang \& Lai (2009b) showed that these modes,
which consist of nearly horizontal oscillations with no vertical
structure, can grow in amplitude due to wave absorption at the
corotation resonance (where the wave pattern speed $\omega/m$
matches the disc rotation rate $\Omega$). This overstability
requires that the disc vortensity, $\zeta \equiv
\kappa^2/(2\Omega\Sigma)$, where $\kappa$ is the radial epicyclic
frequency and $\Sigma$ is the surface density (see Horak \& Lai
2013 for the full general relativistic version of vortensity),
have a positive gradient at the corotation radius $r_c$ (Tsang \&
Lai 2008; see also Narayan et al.~1987). General relativity (GR)
plays an important role: For a Newtonian disc, with
$\Omega=\kappa\propto r^{-3/2}$ and a relatively flat $\Sigma(r)$
profile, $d\zeta/dr<0$, so corotational wave absorption leads to
mode damping; with GR, however, $\kappa$ is non-monotonic in the
inner disc (e.g., for a Schwarzschild BH, $\kappa$ reaches a
maximum at $r = 8GM/c^2$ and goes to zero at $r_{\rm isco} =
6GM/c^2$), thus p-modes with frequencies such that $d\zeta/dr>0$
at $r_c$ are overstable. Linear calculations based on
pseudo-Newtonian potential (Lai \& Tsang 2009; Tsang \& Lai 2009b)
and on full GR (Horak \& Lai 2013) show that the mode frequencies
approximately agree with the observed HFQPO frequencies (with the
BH mass and spin as constrained/measured by observations),
although for rapidly rotating BHs, the linear mode frequencies are
somewhat too large (Horak \& Lai 2013; see below). Nonlinear
simulations show that these overstable modes can grow to the large
amplitudes and maintain global coherence and well-defined
frequencies (Fu \& Lai 2013).  The mode growth and saturation can
also be enhanced by turbulent viscosity (Miranda, Horak \& Lai
2015). Overall, these hydrodynamical studies show that
non-axisymmetric disc p-modes is a promising candidate to explain
HFQPOs in BH X-ray binaries, although a robust diagnostics remains
out of reach due to the complexity of the real systems.

Magnetic fields are likely present in BH accretion discs. They may
be created as a result of the nonlinear development of the
magneto-rotational instability (MRI; Balbus \& Hawley 1998).
Large-scale poloidal fields may be advected inward with the
accretion flow, building up significant strength in the inner disc
(e.g., Lubow et al.~1994; Lovelace et al.~2009; Guilet \& Ogilvie
2012,2013; Cao \& Spruit 2013).  Such large-scale fields can lead
to production of jets/outflows from accretion discs through the
magneto-centrifugal mechanism (e.g., Blandford \& Payne 1982).
Magnetic fields threading the BH can also lead to relativistic
jets from the rotating BHs (McKinney et al. 2012).  As noted
before, in BH X-ray binaries, episodic jets are observed in the
intermediate state, and this is the same state during which HFQPOs
are detected. This suggests that to properly study the BH disc
oscillation modes, it is necessary to include the effect of
large-scale magnetic fields.
%
%
In our model, the disc and corona (coupled by a
large-scale poloidal magnetic field) oscillate together.
We suggest that in the intermediate state, large-scale magnetic
fields are created (perhaps episodically;
cf. Yuan et al.~2009). This would allow
the production of episodic jets and the disk oscillations to
manifest as HFQPOs in hard X-rays as observed (Remillard \& McClintock 2006;
Belloni et al.~2012).
Tagger and collaborators (Tagger \& Pellat 1999;
Varniere \& Tagger 2002; Tagger \& Varniere 2006) have developed a similar
picture of disc oscillations, which they termed accretion-ejection instablity,
although they focused on the MHD form of Rossby wave instability in the disc
(see Lovelace et al. 1999; Yu \& Li 2009; Yu \& Lai 2013).
We note that coherent toroidal magnetic fields in the disc
tend to suppress the corotational instability (Fu \& Lai 2011).
We will focus on poloidal field configurations in this paper.



Recent study by Horak \& Lai (2013) provides a full GR corotation
instability criterion for disc p-modes.
Their full GR results of the mode frequencies are
qualitatively in agreement with the pseudo-Newtonian results (Lai \&
Tsang 2009), but indicate that the theoretical frequencies are too
high compared to observations.
The discrepancy is most severe for rapidly spinning BHs (such as GRS 1905+105).
We show in this paper that including large-scale poloidal fields in the disc
may resolve this discrepency.

In our model, we consider fluid perturbations that have no
vertical structure inside the disc (i.e., the vertical wavenumber $k_z = 0$).
Thus we do not include MRI, which generally involves perturbations with finite
$k_z$. Since the growth of the p-modes is primarily due to in-disc motion,
it is adequate to consider 2D disc dynamics.
Similar setup has previously been considered by various authors in
different contexts (e.g. Spruit et al. 1995; Tagger \& Pallet
1999; Lizano et al. 2010).


Our paper is organized as follows. In \S 2, the basic disc
equations with large-scale magnetic fields are derived. In \S 3,
we present the results of p-modes in thin magnetized discs. In \S
4, we consider the effects of finite disc thickness. and we
conclude in \S 5.

\section{Basic Equations}



We consider a geometrically thin disc and adopt cylindrical
coordinate system $(r,\phi,z)$. We use the the pseudo-Newtonian
potential of Paczynski \& Wiita (1980)
\begin{equation}
\Phi = - \frac{GM}{r-r_s} \ ,
\end{equation}
where $r_s = 2GM/c^2$ is the Schwarzschild radius. For this potential,
the free particle (Keplerian) orbital frequency is
\begin{equation}
\Omega_K = \sqrt{\frac{GM}{r}}\frac{1}{r-r_s},
\end{equation}
and the radial epicyclic frequency $\kappa$ is given by
\begin{equation}
\kappa = \Omega_K \sqrt{r-3r_s\over r-r_s}.
\end{equation}
The innermost stable circular orbit (ISCO), defined
by $\kappa^2=0$, is located at $r_{\rm isco}=3r_s$.
The epicyclic frequency reaches a peak
at $r_{\rm max}=(2+\sqrt{3})r_s$.
Note that the GR effect plays an essential role in
the corotational instability of $p-$modes. In Newtonian theory,
$\kappa = \Omega\propto r^{-3/2}$, so $d\zeta/dr<0$ if the surface density
$\Sigma$ is constant; with GR, $d\zeta/dr>0$ for $r<r_{\rm max}$.

When large-scale magnetic fields thread the thin conducting disc,
the height-integrated mass continuity equation and momentum
equation read
\ba && \frac{\partial\Sigma}{\partial t}
+ \nabla_\perp\cdot\left( \Sigma \mathbf{u} \right) =0,\\
&& \frac{d \mathbf{u}}{dt} = -\frac{1}{\Sigma}\nabla_\perp P +
\frac{1}{4\pi\Sigma} B_z \left[ \mathbf{B} \right]^{+}_{-} +
\mathbf{g},
\label{momentumeqn} \ea
where $\nabla_\perp$ is 2D operator
acting on the disc plane, $\Sigma$, $P$ and ${\bf u}$ are the
surface density, height-integrated pressure and height-averaged
velocity, respectively, $\mathbf{g}=-g(r){\hat r}$, with
$g=GM/(r-r_s)^2$, is the gravitational acceleration, $[{\bf
B}]^+_-\equiv {\bf B}(z=H)-{\bf B}(z=-H)$ (with $H$ the
half-thickness of the disc), and we have used $[B^2]^+_-=0$.
The effect of finite disc thickness will be considered in Section 4.

The unperturbed velocity of an equilibrium disc is ${\bf
u}=(0,r\Omega,0)$ (in cylindrical coordinates), with the angular
velocity given by
\begin{equation}
- \Omega^2 r = - \frac{1}{\Sigma}\frac{dP}{dr} - \frac{d \Phi}{d
r} + \frac{B_z}{2\pi\Sigma}B_r^{+}, \label{eq:equil}
\end{equation}
where $B_r^+=B_r(z=H)=-B_r^-$. The unperturbed disc has
$B_\phi^+=0$. Note that $B_r$ is nonzero only outside the discs,
which has different signs above and below the disc. Inside
the disk, $B_r$ is zero (so that the differential rotation of the disc
would not lead to generation of $B_{\phi}$ inside the disc).
{\bf It should be mentioned that the field configuration adopted in
this paper is highly idealized. A magnetocentrifugal disc wind
or jet, if present, will certainly involve $B_{\phi}$ in the background state.
For simplicity, we assume that the magnetic field outside the disc to be
a potential field.}
The disc surface density is assumed to have a power law form $
\Sigma \propto r^{-p} $, where $p$ is the density index. {\bf We
also try various behavior of the surface density and the magnetic
field, in which the profile of  $B_z/\Sigma$ can be increasing,
decreasing, or constant. We find that the results are
qualitatively similar. Without loss of generality, we fix $p=1$
throughout this paper.  }

The dynamics of the disc is coupled with the large scale
poloidal magnetic field outside the disc, i.e., the disc
magnetosphere. We now consider small-amplitude perturbations of the
disc. We assume that all perturbed quantities have the dependence
$\exp(im\phi-i\omega t)$, where $m=1,2,\cdots $ is the azimuthal
wave number and $\omega$ is the complex frequency. Then the
linearized fluid perturbation equations become
%
\ba && - i \tilde{\omega}\ \delta\Sigma = -\nabla_\perp\cdot
\left(\Sigma\
\delta \mathbf{u}\right),\label{eq:dsigma}\\
&& - i \tilde{\omega}\ \delta u_r - 2\Omega\ \delta u_{\phi} =
-\frac{\partial}{\partial r}\ \delta h
+ \frac{B_z}{2\pi\Sigma} \ \delta B_r^+ \nonumber\\
&&\qquad\qquad\qquad\qquad\quad
+ \frac{B_r^+}{2\pi} \delta\left(\frac{B_z}{\Sigma} \right),\\
&&- i \tilde{\omega}\ \delta u_{\phi}+ \frac{\kappa^2}{2\Omega}\
\delta u_r = - \frac{i m}{r}\ \delta h +\frac{B_z}{2\pi\Sigma} \
\delta B_\phi^+,
\ea
where $\tilde{\omega} = \omega - m\Omega$ is the wave frequency in the
rotating frame of the fluid,
and $\delta h=\delta P/\Sigma$ is the enthalpy
perturbation (we assume barotropic discs). The magnetic field
induction equation in the disc reads
\begin{equation}
- i \tilde{\omega}\ \delta B_z = -  \nabla_\perp \cdot \left( B_z\
\delta\mathbf{u}  \right).
\label{eq:dbz}\end{equation}
Combining Eqs.~(\ref{eq:dsigma}) and (\ref{eq:dbz}), we have
\be
\delta \left({B_z\over\Sigma}\right)=-\xi_r
{d\over dr}\left({B_z\over\Sigma}\right),
\label{eq:dbsigma}\ee
where $\xi_r=\delta u_r/(-i\tilde\omega)$
is the Lagrangian displacement. Equation (\ref{eq:dbsigma})
can also be derived from
$(d/dt)(B_z/\Sigma)=0$.
In terms of $\xi_r$ and $\delta h$, the magnetic field perturbation is
\be
\delta B_{z} = D_1 \xi_r + D_2 \delta h,
\label{eq:dbz0}\ee
with
\be
D_1=B_z{d\over dr}\left(\ln{\Sigma\over B_z}\right),
\quad \ D_2 = \frac{B_z}{c_s^2},
\ee
where $c_s$ is the disc sound speed.

%

To determine $\delta B_r^+$ and $\delta B_\phi^+$, we assume that
the magnetic field outside the disc is a potential field (see
Spruit et al.~1995; Tagger \& Pallet 1999).
This is equivalent to assuming that the Alfven speed above and below the
disc is sufficiently high that currents are dissipated rapidly (on
the disc dynamical timescale). Define the ``magnetic potential''
$\delta\Phi_M$ outside the disc via
\be \delta{\bf B}=-{\rm
sign}(z)\nabla\delta\Phi_M. \ee
Then $\delta\Phi_M$ satisfies the
Poisson equation (Tagger \& Pallet 1999)
\be \nabla^2
\delta\Phi_{M} = -2\,\delta B_z\,\delta(z), \label{poisson} \ee
where $\delta(z)$ is the Dirac delta function.
The integral solution of (\ref{poisson}) is
\begin{equation}
\delta \Phi_{M}(r) = \int \delta B_z (r^{\prime})\,
\left[ \frac{\alpha}{2} \ b^{m}_{1/2}( \alpha ) \right]
dr^{\prime},
\label{lapcoeff1}
\end{equation}
where $\alpha = r^{\prime}/r$, and
the Laplace coefficient is defined by
\begin{equation}
b_s^{m}(\alpha) = \frac{2}{\pi} \int^{\pi}_0 \frac{\cos
m\phi}{(\alpha^2 + \epsilon_0^2 + 1 - 2 \alpha \cos\phi)^s}\
d\phi,
\end{equation}
with $\epsilon_0$ the softening parameter
(of order the disc aspect ratio $H/r$).
The perturbed radial magnetic field at the upper disc
surface is
\ba
&&\delta B_r^+ = - \frac{d}{dr}\delta \Phi_{M}\nonumber\\
&&\quad
=\int\!\delta B_z (r^{\prime})
\left(\frac{\alpha}{2r}\right)
\left[ b^{m}_{1/2}(\alpha)
+\alpha \frac{{\rm d}b^{m}_{1/2}(\alpha)}{{\rm d}\alpha}
\right] dr^{\prime},
\label{lapcoeff2}\ea
and the azimuthal field is
$\delta B_{\phi}^+ = -(im/r)\delta \Phi_{M}$.


In our numerical calculations, we use $\xi_r$ and
$\delta h$ as the basic variables. The perturbation equations are
\ba
&&\frac{d \xi_r}{dr}=-\left(\frac{2 m \Omega}
{r\tilde{\omega}} + \frac{1}{r} + \frac{d\ln\Sigma}{dr} \right)\xi_r\nonumber\\
&&\qquad ~~ -\left(\frac{1}{c_s^2} - \frac{m^2}{r^2
\tilde{\omega}^2 } \right) \delta h + \frac{m^2}{r^2
\tilde{\omega}^2} \frac{B_z}{2\pi\Sigma} \ \delta\Phi_M,
\label{eq:dxir}\\
&&\frac{d}{dr} \delta h = \left( \tilde{\omega}^2-\kappa^2 +
\frac{B_r^+}{2\pi\Sigma}D_1 \right) \xi_r +
\frac{2 m \Omega}{r \tilde{\omega}}\ \delta h\nonumber\\
&&\qquad ~~~ - \frac{B_z}{2\pi\Sigma}\frac{d\delta\Phi_{M}}{dr} +
\frac{2 m \Omega}{r \tilde{\omega}} \frac{B_z}{2\pi\Sigma}\
\delta\Phi_{M}. \label{eq:dh}
\ea

To obtain global trapped modes, wave reflection must occur at the
inner disc boundary $r_{\mathrm{in}}$. To focus on the effect of
corotational instability,
we adopt a simple inner boundary
condition, i.e., the radial velocity perturbation vanishes at the
inner boundary, $\delta u_r = 0$.
{\bf This implies a reflecting inner disc edge, for example
due to the presence of a magnetosphere around the black hole
(e.g., Bisnovatyi-Kogan \& Ruzmaikin 1974, 1976;
Igumenshchev et al. 2003; Rothstein \& Lovelace 2008;
McKinney et al.~2012; see
Tsang \& Lai 2009a and Fu \& Lai 2012 for more detailed treatment of
the magnetosphere-disc interface oscillations).}
As we are interested in the self-excited modes in the inner region of the
disc, we implement the radiative outer boundary conditions such
that waves propagate away from the disc (e.g., Yu \& Li 2009). The
ordinary equations (\ref{eq:dxir}) and (\ref{eq:dh}), together
with the two boundary conditions at the inner and outer disk edge,
form an eigenvalue problem.
Since the eigenfrequency $\omega$ is in general complex, equations
(\ref{eq:dxir})-(\ref{eq:dh}) are a pair of first-order
differential equations with complex coefficients which are
functions of $r$. We solve these equations using the relaxation
method (Press et al.~1992), replacing the ODEs by
finite-difference equations on a mesh of points covering the
domain of interest (typically
$r_{\mathrm{isco}}<r<2.5r_{\mathrm{isco}}$).
According to Eqs.~(\ref{eq:dbz0}), (\ref{lapcoeff1})
and (\ref{lapcoeff2}), both the magnetic potential and its derivative
can be expressed in terms of the linear combination of $\xi_r$ and
$\delta h$. For numerical convenience, we calculate the terms in the square
bracket in Eqs.~(\ref{lapcoeff1}) and (\ref{lapcoeff2}) and store
them for later use. Note that these terms are computed only once
and can be used repetitively. The wave equations (\ref{eq:dxir})
and (\ref{eq:dh}) can be cast in a matrix form that only deals
with the variables $d\xi_r$ and $dh$. Standard relaxation scheme
can be applied to the resulting matrix.
We use uniform grid points in our calculations. The grid point
number is typically chosen to be 350. The relaxation method
requires an initial trial solution that is improved by the
Newton$-$Raphson scheme.
After iterations the initial trial solution converges to the
eigenfunction of the the two-point boundary eigenvalue problem.

\section{Results for Thin Discs}

\begin{figure}
 \includegraphics[width=84mm]{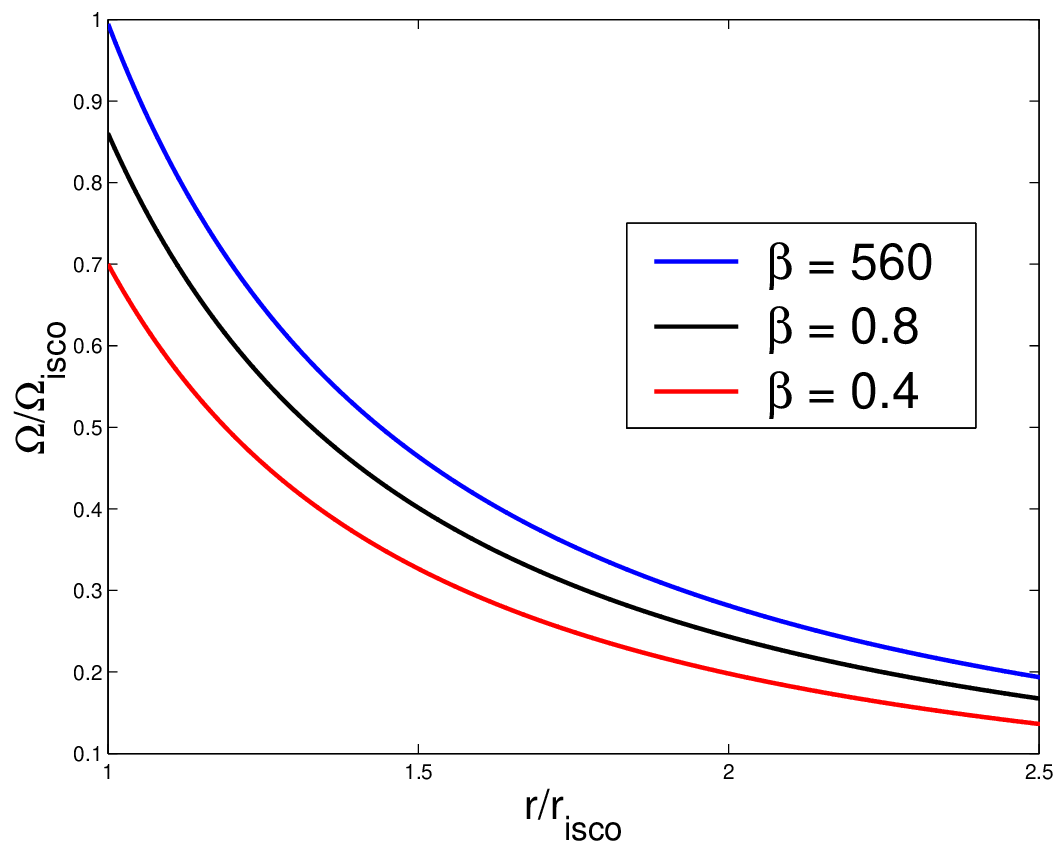}
 \caption{\label{fig00}
  Rotation curve for different value of $\beta$ when $B_r^+ = B_z$.
  With the increase of
  magnetic stress, the angular velocity is reduced.}
\end{figure}

For our numerical calculations, we adopt the disc sound speed $c_s = 0.1 (r
\Omega_K)$. The magnitude of $B_z$ is specified by the
dimensionless ratio
\be \hat B_z={B_z\over
(\Sigma_0\Omega_0^2r_0)^{1/2}}, \ee
where the subscript ``0''
implies that the quantities evaluated at $r=r_{\rm{isco}}$.
The corresponding plasma $\beta$ parameter in the disc at
$r=r_{\mathrm{isco}}$ is \be \beta_0 = \frac{8\pi \rho
c_s^2}{B_z^2}\bigg|_{r=r_{\mathrm{isco}}}={4\pi H_0\over r_0}
\hat B_{z0}^{-2}, \label{eq:beta} \ee where we have used $H =
c_s/\Omega_K\simeq c_s/\Omega$. The magnetic field is chosen such
the plasma $\beta$ is equal to $\beta_0$ at $r=r_{\rm{isco}}$
throughout the disc. Note that magnetic field $B_z$ in the disc
varies approximately as $B_z \propto r^{-1-p/2}$, where $p$ is the
surface density power law index.
We solve for the equilibrium rotation profile using
Eq.~(\ref{eq:equil}). {\bf Note that when $B_r^+=0$, the rotation
profile of the disc is unaffected by $B_z$. When $B_r^+=B_z$, the
equilibrium rotation profile of the disc is changed from the
nonmagnetic disc. In Fig.~\ref{fig00} we show the rotation curve
in the case of $B_r^+ = B_z$ for different value of $\beta$. We
clearly see the reduction in the angular velocity of the
equilibrium disc due to the outward $B_z B_r^+$ stress on the
disc. }

\begin{figure}
 \includegraphics[width=84mm]{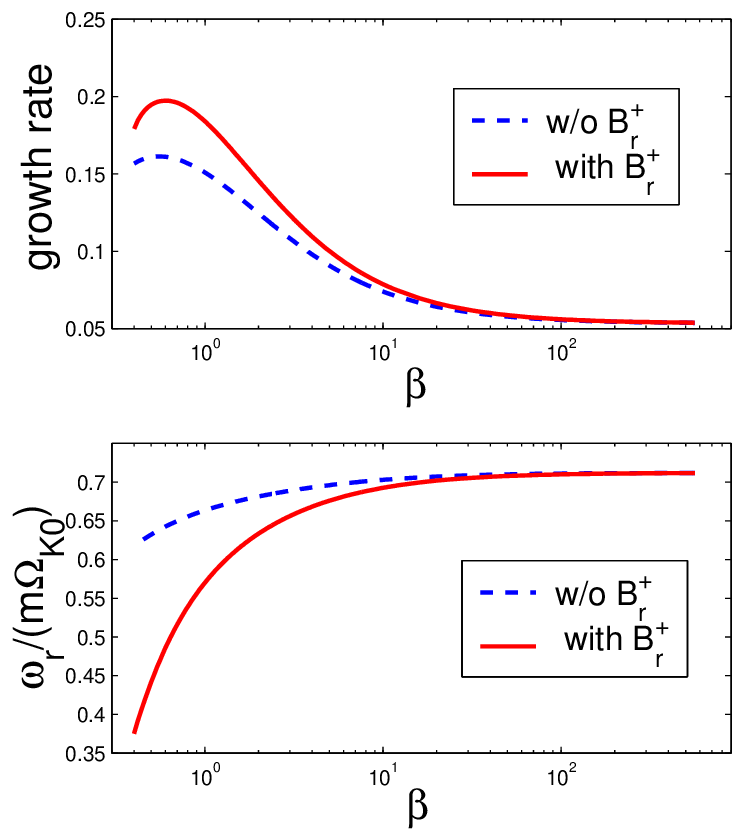}
 \caption{\label{fig1}
The growth rate (in units of $\Omega_{K0}=\Omega_{\rm isco}$)
 (upper panel) and
the real frequency (lower panel) of the $m=2$ p-mode as a function
of the plasma $\beta$ parameter for thin discs [with the
corresponding vertical magnetic field strength $B_z$ given by
Eq.~(\ref{eq:beta})]. The blue dashed line is the case of $B_r^+ =
0$, and the red solid line $B_r^+=B_z$. }
\end{figure}

\begin{figure*}
   \vspace{1mm}
   \begin{center}
   \hspace{5mm}\psfig{figure=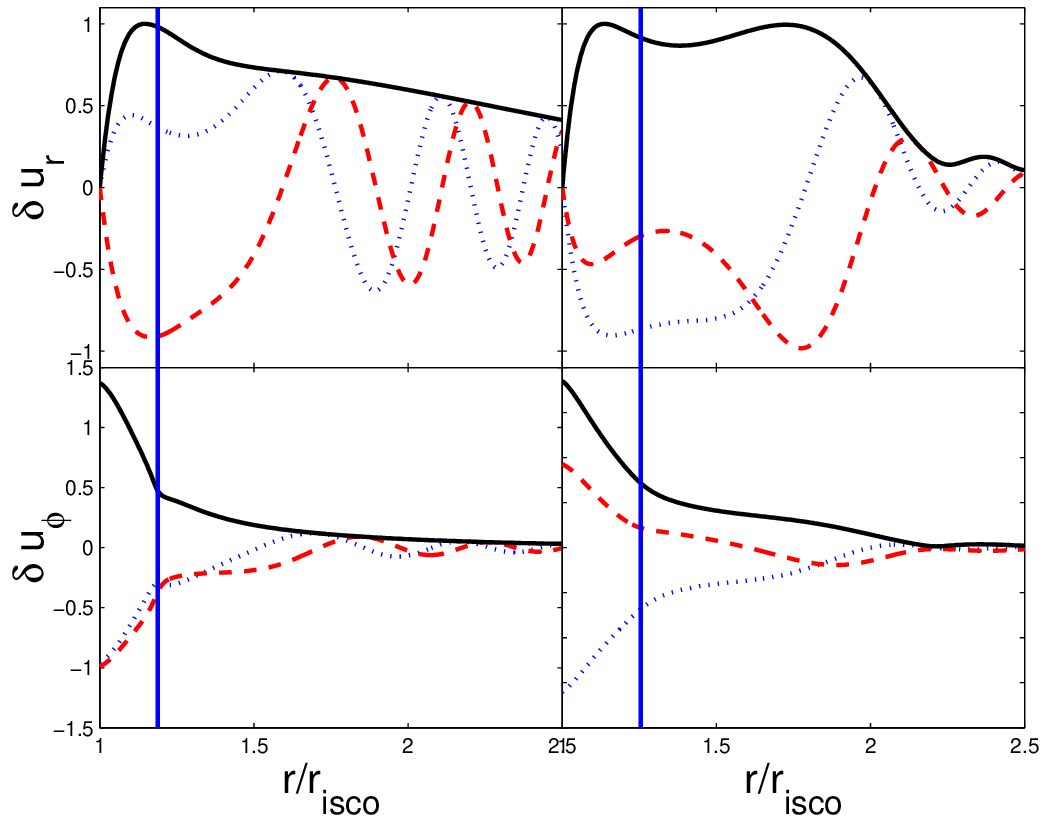,width=160mm,height=90mm,angle=0.0}
   \parbox{100mm}{{\vspace{6mm} }}
   \caption{
   \label{fig3a}
Eigenfunctions of the $m=2$ the p-modes trapped in the inner most
region in the disc. The left panels show the weakly magnetized
case ($\beta \sim 600$); the difference with the nonmagnetic case
is essentially indistinguishable. The right panels have $\beta
\sim 0.6$ (strongly magnetized). The top panels show $\delta u_r$
(the dotted line is for the real part, the dashed line the
imaginary part, and the solid line the absolute value), and the
bottom panels show $\delta u_{\phi}$.
{\bf The corotation resonances
are marked by the vertical lines, which are located at $r_c =1.19, 1.26$,
respectively. }}
   \end{center}
\end{figure*}

\begin{figure}
 \includegraphics[width=84mm]{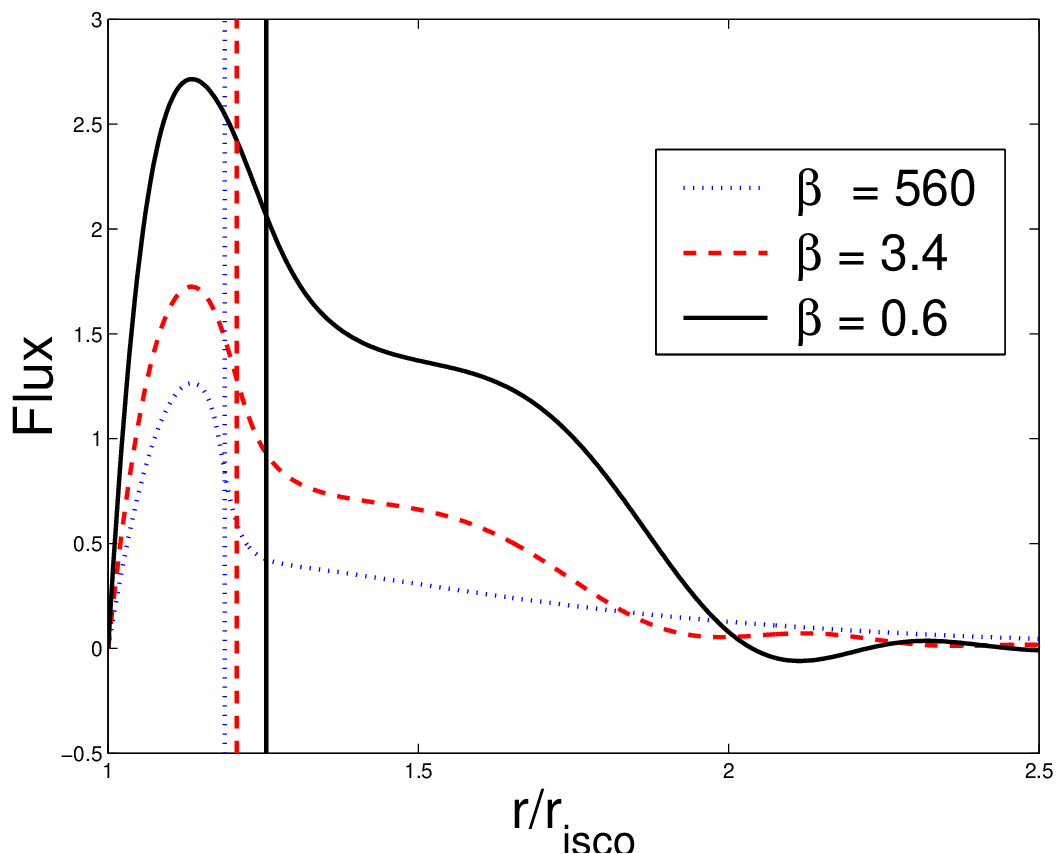}
 \caption{\label{fig3b}
The angular momentum flux associated with the p-mode in disc with
${\hat B}_r^+ = 0$. The blue dotted line is for the weakly
magnetized disc ($\beta \sim 600$), and the red dashed line and
black solid line are for more strongly magnetized disc ($\beta =
3.4 $ and $0.6$). In all cases, the eigenfunction is normalized so
that the maximum velocity perturbation, $\delta u_r$, equals
unity. The higher angular momentum flux for the more strongly
magnetized case is consistent with the larger mode growth rate.
{\bf The vertical lines denote the corotation resonances,
which are located at $r_c = 1.19, 1.21, 1.26$, respectively.}
 }
\end{figure}

The upper panel of Figure 2 shows the growth rate of the p-modes
for different magnetic field strengths. When $B_r^+ = 0$, we see
that with the increase of magnetic field, the p-mode growth rate
varies in a non-monotonic way. The growth rate first increases and
then decreases. At $\beta \sim 0.5$,
the growth rate is enhanced by a factor of 3 compared to the
$B_z=0$ value.  When $B_r^+=B_z$, the overall behavior of the
growth rate is similar to the case of $B_r^+=0$, but the maximum
growth rate is about 30\% higher than disc with $B_r^+=0$. The
mode frequencies are shown in the lower panel in Fig.~2. For the
disc with $B_r^+ = 0$, the mode frequency decreases monotonically
from $\omega_r \simeq 0.7 m\Omega_{K0}$ to $\omega_r \simeq 0.6
m\Omega_{K0}$ with the increase of magnetic field. For disc with
$B_r^+ = B_z$, the frequency can be reduced by a factor about 2,
changing from $\omega_r \simeq 0.7 m\Omega_{K0}$ to $\omega_r
\simeq 0.35 m\Omega_{K0}$.




Figure \ref{fig3a} gives some examples of the $m=2$ eigenfunctions
of p-modes trapped in the innermost region of the disc. The left
panels show
the case with very weak magnetic fields,
and the the right panels show
the magnetized case with $\hat B_r^+ = \hat B_z = 0.78$.
The amplitudes of the eigenfunctions are normalized so that the
maximum absolute value of the radial velocity perturbation $|\delta u_r|$
equals unity (this maximum occurs at $r\simeq r_0$).


To understand the origin of the enhanced p-mode growth rate for
magnetized discs, we show in Fig.~\ref{fig3b} the angular momentum
flux associated with the eigenmode for the weakly magnetized disc
(plasma $\beta \sim 600$) and for the more strongly magnetized
discs (plasma $\beta = 3.4$ and $0.6$). The angular momentum flux
$F(r)$ across a cylinder of radius $r$ is given by (e.g.,
Goldreich \& Tremaine 1979)
\begin{equation}
F(r) =\left\langle r^2 \int^{2\pi}_0 \Sigma \delta u_r \delta
u_{\phi} d\phi \right\rangle = \pi \Sigma r^2 \Re \left( \delta
u_r \delta u_{\phi}^{*} \right)
\end{equation}
where $\langle\rangle$ designates time average and the superscript
$*$ denotes complex conjugate. Note that waves carry negative
(positive) angular momentum inside (outside) the corotation. The
net positive (outward) angular momentum flux $F(r)$ around the
corotation indicates the growth of the p-modes. Higher angular
momentum fluxes imply higher instability growth rates (see Fig.
\ref{fig3b}).


\section{Effects of Finite Disc Thickness}

\subsection{Model Equations}

When finite thickness of the disc is considered, an ``internal''
magnetic force term should be added to the right-hand side of
Eq.~(\ref{momentumeqn}):
\be \mathbf{f} =
\frac{1}{\Sigma}\int\!dz\, \left[ - \nabla_\perp
\left(\frac{B^2}{8\pi} \right) + \frac{1}{4\pi} \left( \mathbf{B}
\cdot \nabla_\perp \right) \mathbf{B} \right]. \ee
Obviously, to
include the 3D effect rigorously would require examining the
vertical stratification of the density and magnetic field inside
the disc -- this is beyond the scope of this paper. Here we
consider a simple model where the internal density of the disc is
assumed to be independent of $z$, so that \be \Sigma = 2\rho
H,\quad \ P = 2 p H. \ee Then the internal magnetic force
simplifies to %
\be %
\mathbf{f} = \frac{1}{4\pi\rho} \left[-\nabla_\perp
\left(\frac{B^2}{2}\right) + \left( \mathbf{B}\cdot
\nabla_\perp\right)\mathbf{B}\right]. %
\ee %
We assume that only vertical magnetic field exists inside the unperturbed
disc. The equilibrium rotational profile is then determined by
\begin{equation}\label{eq:equil3D}
- \Omega^2 r = - \frac{1}{\Sigma}\frac{dP}{dr} - g + %
\frac{B_z}{2\pi\Sigma}B_r^{+} %
- \frac{B_z}{2\pi\Sigma}\left(H \frac{d B_z}{dr} \right) \ . %
\end{equation}
To derive the modified perturbation equations including $\delta {\bf f}$,
it is convenient to define a new perturbation variable
$\delta\Pi$ in place of $\delta h$:
\begin{equation}
\delta \Pi \equiv \frac{\delta P}{\Sigma} + \frac{B_z\delta
B_z}{4\pi\rho} = c_s^2 \frac{\delta \rho}{\rho} + \frac{B_z\delta
B_z}{4\pi\rho}.
\end{equation}
After some algebra, the final disc perturbation equations
can be written in the following form:
\ba
&&\frac{d\xi_r}{dr}=-\left(\frac{2m\Omega}{r\tilde{\omega}} +
\frac{1}{r} + \frac{d\ln\Sigma}{dr} + D_4 \right) \xi_r \nonumber\\
&&\qquad\quad
+\left(\frac{m^2}{r^2\tilde{\omega}^2} - D_5\right)\delta\Pi
+ \frac{m^2}{r^2\tilde{\omega}^2}\frac{B_z}{2\pi\Sigma}\delta\Phi_{M},\\
&&\frac{d}{dr}\delta\Pi = \left(\tilde{\omega}^2 - \kappa^2
+\frac{B_r^+}{2\pi\Sigma}D_3
+ D_6 \right) \xi_r \nonumber\\
&&\quad\qquad + \left(\frac{2m\Omega}{r\tilde{\omega}}
+ D_7 - \frac{d\ln\rho}{dr}\right)\delta \Pi\nonumber\\
&&\qquad\quad +\frac{2m\Omega}{r\tilde{\omega}} \frac{B_z}{2\pi\Sigma}
\delta\Phi_{M} - \frac{B_z}{2\pi\Sigma}\,\frac{d\delta\Phi_{M}}{dr},
\ea
where $D_3=D_1$, and
\ba
&& D_4 =\frac{-\ c_a^2}{c_s^2 + c_a^2}
{d\over dr}\left(\ln {\Sigma\over B_z}\right),\quad
c_a^2 = \frac{B_z^2}{4\pi\rho},\\
&&
D_5 = \frac{1}{c_s^2 + c_a^2},\\
&& D_6 = \left[
\frac{1}{\rho}\frac{d(p+\frac{1}{8\pi} B_z^2)}{dr} \right] D_4,\\
&&D_7 = \left[ \frac{1}{\rho}\frac{d(p+\frac{1}{8\pi} B_z^2)}{dr}
\right] D_5.
\ea

\subsection{Results}

\begin{figure}
 \includegraphics[width=84mm]{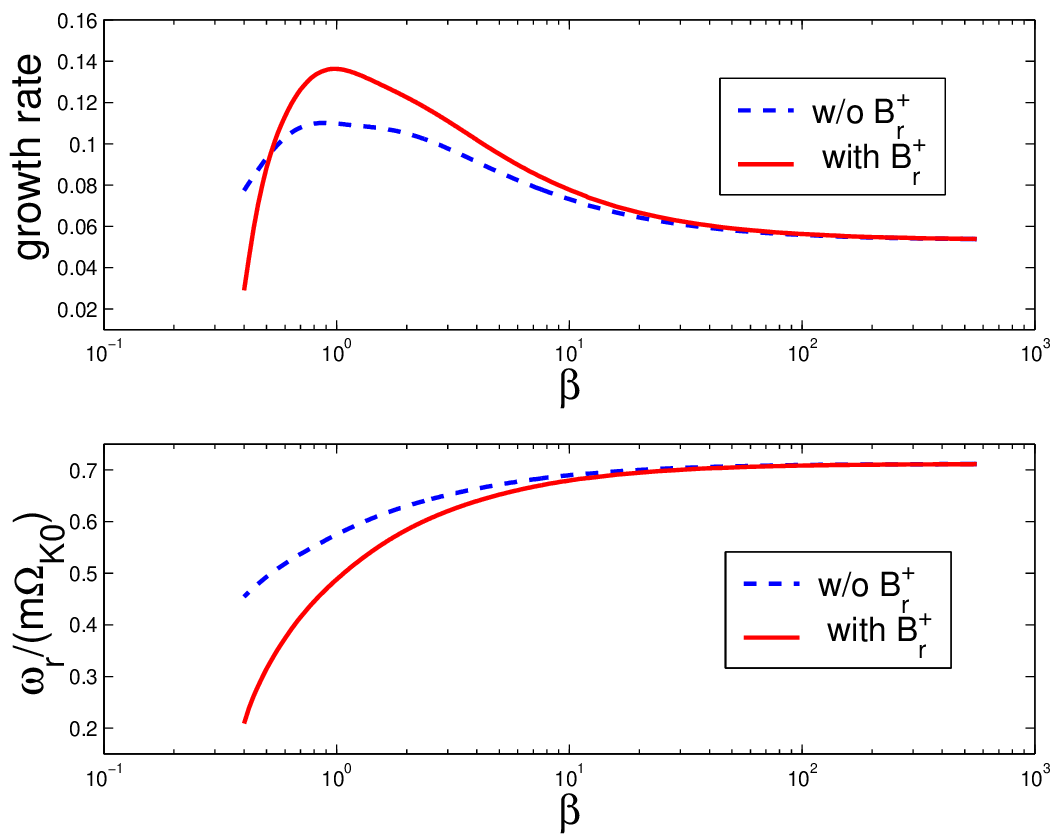}
 \caption{\label{fig4}
Growth rate (in units of $\Omega_{\rm K0}=\Omega_{\rm isco}$) and
frequency of the unstable $m=2$ p-mode as a function of the plasma
$\beta$ parameter for discs with finite thickness. The blue dashed
line is for the case with $B_r^+=0$, while the red solid line for
the case of $B_r^+ = B_z$.
 }
\end{figure}

Finite disc thickness has two-fold effects on the p-mode
corotation instability : the first is the change in the equilibrium
rotation profile and the second is the direct effect of $\delta
\mathbf{f}$.
With these two effects included, we calculate the eigenmodes of
the disc using the same method as in Section 3. The results are
shown in Figure \ref{fig4}. The red solid line shows the case with
$B_r^+ = B_z$ and the blue dashed line shows the case with $B_r^+
= 0$. We find that for the parameter considered ($c_s=0.1
r\Omega_K $, or $H/r=0.1$), the p-mode growth rate is decreased
compared to the $H/r\rightarrow 0$ limit.
More specifically, the maximum growth rate is about $25\%-30\%$
lower. However for sufficiently strongly-magnetized disc (plasma
$\beta \sim 0.4$), especially for disc with radial magnetic field
in the magnetosphere, the oscillation frequency of the overstable
mode can be reduced further, changing from $\omega_r \simeq 0.7
m\Omega_{K0}$ to $\omega_r \simeq 0.2 m\Omega_{K0}$, which is
reduced by about of factor 3 compared to the pure hydrodynamic
mode.

As noted in Section 1, full GR calculations of hydrodynamic disc
p-modes show that, for accretion discs around extreme Kerr BHs,
the oscillation frequency is close to $m\Omega_{K0}$ (Horak \& Lai
2013). However, observations reveal that the frequencies of HFQPOs
are usually smaller than this value by a factor of 2-3. This
discrepancy indicates that pure hydrodynamic disc p-modes without
magnetic field can not explain HFQPOs in X-ray binaries,
particularly for high-spin sources (such as GRS 1905+105). When
the large-scale poloidal magnetic field is included, this
discrepancy can be resolved since the reduction in oscillation
frequency makes the theoretical p-mode prediction more consistent
with the observed values.

In addition, we find that the first effect, i.e., the change of
the equilibrium disc rotation profile {\bf induced by the disk
finite thickness}, plays a minor role in the disc p-mode
instability. To make this point more clearly, we artificially
discard the last term on the right hand side of Equation
(\ref{eq:equil3D}). We find that results are very close to those
with this term included (the detailed results are not shown in
Fig. 5, since they are very close to the curves in Fig.~5).

\section{Discussion and Conclusion}


In this paper we have carried out linear analysis of p-modes
in BH accretion discs threaded by large-scale magnetic fields. These modes
reside in the innermost region in the disc, can become overstable
driven by corotational instability,
and may be responsible for the observed high-frequency
quasi-periodic oscillations (HFQPOs) in BH X-ray binaries.
Our results show that the large-scale magnetic field can
increase the mode growth rates significantly (by a factor of
a few for plasma $\beta\sim 1$ in the disc). It can also reduce
the mode frequencies, bringing the theoretical values into
agreement with the observed HFQPO frequencies.

It is important to note that since episodic jets are observed in the
intermediate state of BH X-ray binaries when HFQPOs are detected,
a proper treatment of large-scale magnetic fields is crucial
for understanding the disc-jet-QPO connections in accreting BH systems
(e.g. Fender et al.~2004).
Although our treatment of the disc magnetic fields is somewhat idealized
(e.g., toroidal fields, vertical stratification and
MRI turbulence are not included), it demonstrates the importance of
including magnetic fields in studying BH disc oscillations
and confronting with observations.

Overall, our study shows that global oscillations of BH accretion
discs, combining the effects of general relativity and magnetic
fields, is a promising candidate for understanding HFQPOs observed
in BH X-ray binaries and similar oscillations potentially observed
in intermediate and supermassive BHs
(e.g., Pasham, Strohmayer \& Mushotzky 2014). Important caveats
and uncertainties (see Lai et al.~2013 for a concise review of
various issues) remain in our treatment of the physical condition
of the innermost accretion flows around BHs (e.g., the BH
magnetosphere - disc interface; see Tsang \& Lai 2009a; Fu \& Lai
2012; Miranda et al.~2015). It is hoped that future observations
with more sensitive X-ray timing telescopes (such as LOFT, see
Feroci et al.~2014) will shed more light on the origin of
variabilities of accreting BHs (e.g. Belloni \& Stella 2014).

\section*{Acknowledgments}
DL thanks Wen Fu, Jiri Horak, Ryan Miranda, David Tsang, Wenfei Yu
and Feng Yuan for useful discussions over the last few years. This
work has been supported in part by NSF grant AST-1008245, 1211061,
and NASA grant NNX12AF85G. CY thanks the support from National
Natural Science Foundation of China (Grants 11173057 and
11373064), Yunnan Natural Science Foundation (Grant 2012FB187,
2014HB048), and Western Light Young Scholar Program of CAS. Part
of the computation is performed at HPC Center, Yunnan
Observatories, CAS, China. Both authors thank the hospitality of
Shanghai Astronomical Observatory, where part of the work was
carried out.


\end{document}